\documentstyle[12pt]{article}

\newcommand{\sect}[1]{\setcounter{equation}{0}\section{#1}}

\def\be{\begin{equation}}
\def\ee{\end{equation}}
\def\ba{\begin{eqnarray}\samepage}
\def\ea{\end{eqnarray}}

\font\twelvemsa=msam10 scaled 1200

\font\sevenmsa=msam7
\font\fivemsa=msam5
\newfam\msafam
\textfont\msafam=\twelvemsa
\scriptfont\msafam=\sevenmsa
\scriptscriptfont\msafam=\fivemsa
\def\msa{\ifcase\msafam 0\or1\or2\or3\or4\or5\or6\or7\or8\or9\or A\or B\or
C\or D\or E\or F\fi}

\font\twelvemsb=msbm10 scaled 1200

\font\sevenmsb=msbm7
\font\fivemsb=msbm5
\newfam\msbfam
\textfont\msbfam=\twelvemsb
\scriptfont\msbfam=\sevenmsb
\scriptscriptfont\msbfam=\fivemsb
\def\msb{\ifcase\msbfam 0\or1\or2\or3\or4\or5\or6\or7\or8\or9\or A\or B\or
C\or D\or E\or F\fi}

\font\twelveeuf=eufm10 scaled 1200

\font\seveneuf=eufm7
\font\fiveeuf=eufm5
\newfam\euffam
\textfont\euffam=\twelveeuf
\scriptfont\euffam=\seveneuf
\scriptscriptfont\euffam=\fiveeuf
\def\euf{\ifcase\euffam 0\or1\or2\or3\or4\or5\or6\or7\or8\or9\or A\or B\or
C\or D\or E\or F\fi}

\def\goth#1{\fam\euffam#1}
\def\Bbb#1{\fam\msbfam#1}
\def\fraction#1#2{{\textstyle\frac{#1}{#2}}}
\def\half{\textstyle\frac{1}{2}}

\mathchardef\gapprox"3\msa26
\mathchardef\lapprox"3\msa2E

\begin{document}

\title{Non-Linear Electrodynamics: Zeroth and First Laws of Black Hole
  Mechanics}

\author{D A Rasheed\\ \\DAMTP\\University of Cambridge\\Silver
Street\\Cambridge\\CB3 9EW}

\maketitle

\begin{abstract}
The Zeroth and First Laws of Black Hole Mechanics are derived in the
context of non-linear electrodynamics coupled to gravity. The Zeroth
Law is shown to hold quite generally even if the Dominant Energy
Condition is violated. The derivation of the First Law is discussed in
detail for general matter fields coupled to gravity. The general mass
variation formula obtained includes a term previously omitted in some
of the literature. This is then applied to the case of non-linear
electrodynamics and the usual First Law is found to hold true. As an
example, Born-Infeld theory is discussed. The results are extended to
include scalar fields in a very general way, including additional
terms arising from the variation of the asymptotic values of the
scalars.
\end{abstract}

PACS: 04.20.-q; 04.40; 04.70; 11.10.Lm

\renewcommand{\thepage}{ }
\pagebreak

\renewcommand{\thepage}{\arabic{page}}
\setcounter{page}{1}

\sect{Introduction}

Theories of non-linear electrodynamics were first proposed in the
1930's as an attempt to construct a classical theory of charged
particles with finite self-energy \cite{BorInf}. In this regard the
theories were successful. However, they did not remove all of the
singularities associated with a point charge and non-linear
electrodynamics became less popular with the introduction of QED which
provided much better agreement with experiment. Now, however,
non-linear electrodynamics is making a comeback as it arises naturally
in open superstring theory and the theory of D-branes.

One non-linear theory in particular, Born-Infeld theory, seems to be
distinguished from all the others, occuring repeatedly in many
different contexts in the study of strings and branes. For example,
loop calculations in open superstring theory lead to a low energy
effective action of the Born-Infeld type \cite{FraTse}. More recent
interest has focussed on type II open superstrings with boundary
states obeying Dirichlet boundary conditions. The resulting D-brane
low energy effective actions, which depend on a world-volume $U(1)$
vector field, are also found to be of the Born-Infeld type \cite{Lei}.
One of the features of Born-Infeld theory is that it imposes a maximum
allowed electric field strength in terms of a fundamental parameter of
the theory. Applied to string theory, this maximum electric field
strength is $1\over 2\pi\alpha'$, where $\alpha'$ is the inverse
string tension. The existence of an upper bound on the the electric
field strength can be interpreted in terms of the pair production of
open strings, with equal and opposite charges at their ends, by a
constant electric field: In the weak field limit the pair production
rate for such open strings coincides with Schwinger's classical result
\cite{Sch}.  However, in \cite{BacPor} it was shown that this rate
diverges as the electric field strength approaches some maximum
critical value of the order of the string tension $1\over\alpha'$.

Born-Infeld electrodynamics also has the remarkable property that,
despite its highly non-linear nature, the equations of motion have an
exact $SO(2)$ electric-magnetic duality invariance \cite{Schro}. By
adding axion and dilaton fields, this invariance may be extended to
$SL(2,{\Bbb R})$ S-duality, relevant to string theory, which implies a
strong-weak coupling duality of such theories. Such dualities were
investigated in the context of more general non-linear electrodynamic
theories \cite{GibRas1,GibRas2} and it was found that there are in
fact an infinite number of such duality invariant theories. However,
Born-Infeld, along with Maxwell theory, is the only one which is known
in an exact closed form.

Recently theories such as the theory of black points, with a
logarithmic electromagnetic Lagrangian coupled to gravity, have been
studied in an attempt to remove some of the singularities associated
with a charged black hole \cite{Sol}. The divergence of the
energy-momentum tensor was successfully removed, although the
spacetime still exhibited a curvature singularity, albeit of a weaker
variety.  Born-Infeld theory has also been investigated with this goal
in mind \cite{Pal}, the results suggesting that the usual point
singularity inside the event horizon is to be replaced by a `ball of
matter' of finite density which somehow resists the squeeze of
gravity.

The exterior of such black holes/points is, at large distances, the
same as the usual black holes of Einstein-Maxwell theory. Close to the
black hole, however, things may be very different. The aim of this
paper is to investigate the thermodynamic properties of such black
holes, focussing mainly on the First Law.

The original derivation of the First Law of Black Hole Mechanics
\be
dM = {\kappa\over 8\pi}d{\cal A} + \Omega_HdJ + \Phi_HdQ
\ee
was done using a covariant approach \cite{BarCarHaw}-\cite{Reviews}.
It applies for perturbations from one stationary axi-symmetric
solution of the Einstein-Maxwell equations to another. Since then
other proofs using the Hamiltonian formalism have been given which are
valid for more general perturbations \cite{Wal}. Covariant techniques
similar to those used in the original derivation of the First Law may
also be used to prove Smarr's formula
\be
M = {\kappa{\cal A}\over 4\pi} + 2\Omega_H J + \Phi_HQ.
\label{Smarr}
\ee
However, it was later realized \cite{GibPer} that, in the case of
Einstein-Maxwell theory, these two formulae are very closely related
and that it is possible to deduce one directly from the other using
the homogeneity of the mass $M$ as a function of $\sqrt{\cal A}$,
$\sqrt{J}$ and $Q$. Thus it is only necessary to derive {\em one} of
the two formulae -- preferably the Smarr formula, since its derivation
is much simpler.

In the case of non-linear electrodynamics, however, one no longer has
homogeneity of the mass function and so, not only is it no longer
possible to pass easily between these two formulae, but also one can
no longer expect both to hold (indeed, a priori one has no reason to
expect that either of them holds!). The main conclusion of this paper
is that the First Law does indeed continue to hold for {\em all}
theories of non-linear electrodynamics (and hence the Smarr formula
does not).

Section~2 contains a brief introduction and description of what is
meant by a non-linear theory of electrodynamics. The conventions being
used will be established and some useful electromagnetic quantities
defined for the case of stationary axi-symmetric black holes. As a
precursor to the main discussion of the First Law, in section~3 some
brief comments will be made concerning the Zeroth Law. The main point
is that, for non-linear electrodynamics, the requirement that the
Dominant Energy Condition be satisfied is no longer necessary to prove
the Zeroth Law. In section~4 the covariant approach will be used to
establish the First Law for a fully general energy-momentum tensor,
describing {\em any} form of matter fields coupled to gravity. The
mass variation formula given will include a term, relevant to rotating
black holes, previously omitted in \cite{HeuStr}. In section~5 this
result will then be applied to non-linear electrodynamics coupled to
gravity. As an example of such a theory, the static black holes of
Born-Infeld theory will be discussed in section~6. Finally, in
section~7, the inclusion of scalar fields is discussed.

Throughout this paper, geometrical units will be used:
$G=c=4\pi\epsilon_0={\mu_0\over 4\pi}=1$.


\sect{Non-Linear Electrodynamics}

The usual (linear) theory of electrodynamics coupled to gravity
(Einstein-Maxwell theory) may be described using the Lagrangian
formulation. The action is
\be
S = \int d^4x \sqrt{g} \left( R - F_{\mu\nu}F^{\mu\nu} \right)
\ee
where $F_{\mu\nu}=\partial_\mu A_\nu-\partial_\nu A_\mu$. Varying with
respect to $A_\mu$ gives the electromagnetic equations of motion
\be
\nabla_\mu F^{\mu\nu} = 0 \quad\hbox{or}\quad d\star F = 0
\ee
which are linear in $F_{\mu\nu}$. This is what is meant {\em linear}
electrodynamics. Note that the Einstein field equations obtained by
varying with respect to the metric $g_{\mu\nu}$ are still of course
non-linear (in both $g_{\mu\nu}$ and $F_{\mu\nu}$):
\be
R_{\mu\nu}-\half Rg_{\mu\nu} = 2\left( {F_\mu}^\lambda F_{\nu\lambda}
  - \fraction{1}{4}g_{\mu\nu}F^2 \right).
\ee
In non-linear electrodynamics one replaces the usual electromagnetic
Lagrangian $L_F=-F^2$ by a more general function of the field strength
$F_{\mu\nu}$. For example, in the theory of black points \cite{Sol}
the Lagrangian $-\log(1+F^2)$ is used. Since the Lagrangian is no
longer quadratic in $F_{\mu\nu}$, the equations of motion are both
non-linear in $F_{\mu\nu}$.

Consider a general theory of non-linear electrodynamics coupled to
gravity, described by the action
\be
S = \int d^4x \sqrt{g} \left( R + L_F \right)
\ee
where $L_F(F_{\mu\nu})$ is assumed to be function of the field
strength tensor $F_{\mu\nu}$ only, not containing any higher
derivative terms in $F_{\mu\nu}$. If the theory is to agree with the
usual Einstein-Maxwell theory for weak fields (as is the case for the
logarithmic Lagrangian above and for Born-Infeld theory which will be
discussed in section~6), then $L_F$ must be of the form
\be
L_F = -F_{\mu\nu}F^{\mu\nu} + {\cal O}(F^4).
\label{TypicalL}
\ee
It is useful to define the second rank tensor $G^{\mu\nu}$ by
\footnote{There is some ambiguity in the definition of this partial
  derivative depending on whether or not one takes into account the
  antisymmetry of $F_{\mu\nu}$.  Here $F_{\mu\nu}$ and $F_{\nu\mu}$
  are treated as independent variables in the partial derivative}
\be
G^{\mu\nu} = -{1\over 2} {\partial L_F\over\partial F_{\mu\nu}}.
\ee
So for Lagrangians of the form (\ref{TypicalL}) $G^{\mu\nu}$ takes
the form $G^{\mu\nu}=F^{\mu\nu}+{\cal O}(F^3)$. Varying the action
with respect to $A_\mu$ gives the electromagnetic field equations
\be
\nabla_\mu G^{\mu\nu} = 0 \quad\hbox{or}\quad d\star G = 0.
\label{EOM}
\ee
Since $G_{\mu\nu}$ is in general a highly non-linear function of
$F_{\mu\nu}$, these equations are non-linear. The field strength
$F_{\mu\nu}$ also satisfies the Bianchi identities
\be
\nabla_{[\mu} F_{\nu\lambda]} = 0 \quad\hbox{or}\quad dF = 0.
\label{Bianchi}
\ee
Note that there is a clear similarity between these last two
equations. If $F_{\mu\nu}$ and $G_{\mu\nu}$ were independent variables
then the theory would have an $SO(2)$ symmetry corresponding to
rotating $F$ into $\star G$ (electric-magnetic duality). However,
since $F_{\mu\nu}$ and $G_{\mu\nu}$ are not independent, this duality
mixing the Bianchi identities with the equations of motion is in
general inconsistent. In the special case of Born-Infeld theory this
duality is an exact symmetry of the theory (see \cite{GibRas1} for a
more complete discussion of electric-magnetic duality in non-linear
electrodynamics). Varying the action with respect to the metric
$g_{\mu\nu}$ gives the Einstein field equations
\be
R_{\mu\nu} - \half Rg_{\mu\nu} = 2T_{\mu\nu}
\ee
where the energy-momentum tensor $T_{\mu\nu}$ for non-linear
electrodynamics is found to be
\be
T_{\mu\nu} = {G_\mu}^\lambda F_{\nu\lambda} +
\fraction{1}{4}g_{\mu\nu}L_F.
\label{EnMom}
\ee
Note that $L_F$ can be a function only of the two scalar invariants
$F^2=F_{\mu\nu}F^{\mu\nu}$ and $F\star
F=\half\epsilon_{\mu\nu\rho\sigma}F^{\mu\nu}F^{\rho\sigma}$ since (in
four dimensions) these are the only two scalar invariants which can be
formed from $F_{\mu\nu}$. This means that $G_{\mu\nu}$ must be of the
form
\be
G_{\mu\nu} = aF_{\mu\nu} + b\star F_{\mu\nu}
\label{Gab}
\ee
for some scalar functions $a$ and $b$. Here $\star$ is the Hodge star
operator given by $\star
F_{\mu\nu}=\half\epsilon_{\mu\nu\rho\sigma}F^{\rho\sigma}$ where
$\epsilon_{\mu\nu\rho\sigma}$ are the components of the covariantly
constant volume 4-form. $F$ and $\star F$ satisfy the useful
identities ${F_\mu}^\lambda\star F_{\nu\lambda}={1\over 4}F\star
Fg_{\mu\nu}$ and $\star{F_\mu}^\lambda\star
F_{\nu\lambda}=-{F_\mu}^\lambda F_{\nu\lambda}$. This ensures that
$T_{\mu\nu}$ as defined above is symmetric.

Consider now stationary axi-symmetric black hole solutions. Let $k$ be
the (unique) time-translation Killing vector, normalized so that
$k^2=-1$ at infinity and let $m$ denote the rotational Killing vector
whose orbits are closed curves with parameter length $2\pi$. Thus the
metric $g_{\mu\nu}$, the field strength tensor $F_{\mu\nu}$ and also
$G_{\mu\nu}$ are invariant under the action of the time-translational
and rotational symmetries generated by $k$ and $m$:
\ba
 & & {\cal L}_kg_{\mu\nu} = {\cal L}_mg_{\mu\nu} = 0 \nonumber\\
 & & \nonumber\\
 & & {\cal L}_kF_{\mu\nu} = {\cal L}_mF_{\mu\nu} = 0 \\
 & & \nonumber\\
 & & {\cal L}_kG_{\mu\nu} = {\cal L}_mG_{\mu\nu} = 0 \nonumber
\ea
where $\cal L$ denotes the Lie derivative. In the usual coordinates
$(t,r,\theta,\phi)$ these Killing vectors are
\be
k = {\partial\over\partial t} \quad\hbox{and}\quad m =
{\partial\over\partial\phi}.
\ee
One may then form a new Killing vector $\xi$ as a linear combination
of $k$ and $m$ which is tangent to the null generators of the black
hole event horizon $\cal H$
\be
\xi = k + \Omega_Hm.
\ee
The requirement that $\xi$ be null on the horizon fixes the constant
$\Omega_H$ which may be interpreted as the angular velocity of the
event horizon.

Electric and magnetic field vectors may now be defined as follows
\be
E_\mu = F_{\mu\nu}\xi^\nu \quad\hbox{and}\quad H_\mu = -\star
G_{\mu\nu}\xi^\nu.
\label{EH}
\ee
The signs are chosen so that in flat space their spatial parts agree
with the usual definitions $E_i=F_{i0}$ and
$H_i=\half\epsilon_{ijk}G_{jk}$.
Using the Bianchi identities (\ref{Bianchi}) one has
\be
\nabla_{[\mu}E_{\nu]} = -\half{\cal L}_\xi F_{\mu\nu} = 0.
\ee
So $E_\mu$ may be written in terms of a (co-rotating) electric
potential $\Phi$ via
\be
E_\mu = \partial_\mu\Phi
\label{EPhi}
\ee
and choosing $\Phi\rightarrow 0$ as $r\rightarrow\infty$ determines it
uniquely.

Note that due to the gauge freedom the vector potential $A_\mu$ is not
necessarily invariant under the stationary and axi-symmetry group
actions (for example, the gauge transformation $A_\mu\rightarrow
A_\mu+\partial_\mu f(t)$ destroys time-translation invariance). So in
general one has
\be
{\cal L}_\xi A_\mu = \Lambda_\mu \ne 0
\ee
where $\Lambda_\mu$ is obviously a gauge dependent quantity. Taking
the exterior derivative of the above equation implies that
$\nabla_{[\mu}\Lambda_{\nu]}=0$ and hence
\be
\Lambda_\mu = \partial_\mu\Lambda.
\ee
Then it can be shown that
\be
E_\mu = \partial_\mu \left( A_\nu\xi^\nu - \Lambda \right)
\ee
so comparison with (\ref{EPhi}) gives
\be
\Phi = A_\nu\xi^\nu - \Lambda + \hbox{const.}
\ee
Typically, for an electrically charged black hole, one would choose a
gauge in which $\Lambda=0$ and $A_\mu\rightarrow 0$ as
$r\rightarrow\infty$. Then {\em in this gauge}
\be
\Phi = A_\nu\xi^\nu
\ee
which is commonly taken as the definition of the co-rotating electric
potential. However, this is not very satisfactory since it is not a
gauge invariant statement. The definition (\ref{EPhi}) is preferable
since it guarantees that $\Phi$ is gauge invariant (at least up to a
constant) and also that it is preserved by the stationary and
axi-symmetry group actions. 

Considering magnetically charged black holes it is also possible to
define a magnetic potential $\Psi$ as follows: Using the
electromagnetic field equations (\ref{EOM}) one has
\be
\nabla_{[\mu}H_{\nu]} = \half{\cal L}_\xi \star G_{\mu\nu} = 0
\ee
and so $H_\mu$ may be written as
\be
H_\mu = \partial_\mu\Psi
\label{HPsi}
\ee
and choosing $\Psi\rightarrow 0$ as $r\rightarrow\infty$ fixes it
uniquely.

Due to the antisymmetry of $F_{\mu\nu}$ and $\star G_{\mu\nu}$,
$E_\mu$ and $H_\mu$ satisfy $\xi.E=\xi.H=0$ and hence they are tangent
to the horizon $\cal H$. Using Raychaudhuri's equation
$R_{\mu\nu}\xi^\mu\xi^\nu=0$ on $\cal H$ and the field equations, one
has $T_{\mu\nu}\xi^\mu\xi^\nu=0$ on $\cal H$. Substituting for
$T_{\mu\nu}$ from (\ref{EnMom}) and using (\ref{Gab}) this implies
that
\be
{F_\mu}^\lambda F_{\nu\lambda} \xi^\mu \xi^\nu = E_\mu E^\mu = 0
\quad\hbox{on } {\cal H}.
\ee
Similarly it is possible to deduce that $H_\mu$ is also null on the
horizon:
\be
H_\mu H^\mu = 0 \quad\hbox{on } {\cal H},
\ee
So $E_\mu$ and $H_\mu$ must be proportional to $\xi_\mu$ on $\cal H$
and hence the potentials $\Phi$ and $\Psi$ take {\em constant} values
$\Phi_H$ and $\Psi_H$ respectively on the horizon.


\sect{Zeroth Law}

The Zeroth Law of Black Hole Mechanics states that the surface gravity
$\kappa$ of a stationary black hole is constant over the event
horizon. The original proof \cite{BarCarHaw,Reviews}, applicable to
any Killing horizon, relies on the energy-momentum tensor satisfying
the Dominant Energy Condition. More recently \cite{KayWal} the Zeroth
Law has been proved without resort to the Dominant Energy Condition,
using only the assumption that the event horizon be a bifurcate
Killing horizon. In this section it will be shown that, in the case of
non-linear electrodynamics, neither of these assumptions is required.

The standard proof of the Zeroth Law proceeds as follows: Define the
vector $J$ by
\be
J_\mu = -T_{\mu\nu}\xi^\nu.
\ee
Then by Raychaudhuri's equation $\xi.J=0$ on $\cal H$, so $J$ is
tangent to the horizon and hence it is spacelike or null there. At
this point one appeals to the Dominant Energy Condition which implies
that $J$ must be timelike or null. Therefore $J$ is null on the
horizon and hence proportional to $\xi$. Now one uses the identity
$\xi_{[\mu}\partial_{\nu]}\kappa=-\xi_{[\mu}R_{\nu]\lambda}\xi^\lambda$
on $\cal H$, which follows from the definition
$\xi^\mu\nabla_\mu\xi^\nu=\kappa\xi^\nu$ on $\cal H$ and the Frobenius
condition $\xi_{[\mu}\nabla_\nu\xi_{\lambda]}=0$ on $\cal H$. This
identity together with the Einstein field equations gives
\be
\xi_{[\mu}\partial_{\nu]}\kappa = 2\xi_{[\mu}J_{\nu]} \quad\hbox{on
  }{\cal H}.
\ee
Now the right hand side vanishes since $J$ is proportional to $\xi$ on
$\cal H$. Hence $\partial_\mu\kappa$ is proportional to $\xi$ which is
normal to the horizon and so $\kappa$ is constant on $\cal H$.

In non-linear electrodynamics, however, the use made of the Dominant
Energy Condition above is unnecessary. Using (\ref{EnMom}) and
(\ref{Gab}) one may write the energy-momentum tensor in the form
\be
T_{\mu\nu} = a{F_\mu}^\lambda F_{\nu\lambda} + \fraction{1}{4}\left(
  L_F + bF\star F \right)g_{\mu\nu}.
\ee
Now, as was proved at the end of the last section, $E_\mu$ is
proportional to $\xi_\mu$ on $\cal H$. Therefore $J_\mu$, as defined
above, must also be proportional to $\xi_\mu$ on $\cal H$, as a direct
consequence of the special form of the energy-momentum tensor for
non-linear electrodynamics. The Zeroth Law then follows as above. Note
that this argument remains true even if the Dominant Energy Condition
is violated and is valid for any type of Killing horizon.


\sect{First Law}

In this section the mass variation formula for stationary black holes
in General Relativity coupled to general matter fields will be derived
in some detail, using the traditional covariant approach. The final
formula obtained will correct a minor error in equation (92) of
\cite{HeuStr} which omits a term proportional to $\delta\Omega_H$.

First some conventions must be established. The Komar integral for the
total mass of a stationary axi-symmetric asymptotically flat spacetime
is given by
\be
M = -{1\over 8\pi} \oint_\infty dS_{\mu\nu} \nabla^\mu k^\nu
\ee
where the integral is over a closed surface, topologically a 2-sphere,
at spatial infinity. Similarly the total angular momentum may be
expressed as a Komar integral:
\be
J = {1\over 16\pi} \oint_\infty dS_{\mu\nu} \nabla^\mu m^\nu.
\ee
It is convenient to define the corresponding integral over the event
horizon which may be interpreted as the horizon angular momentum:
\be
J_H = {1\over 16\pi} \oint_{\cal H} dS_{\mu\nu} \nabla^\mu m^\nu.
\ee
The surface element on the event horizon is given by
$dS_{\mu\nu}=2n_{[\mu}\xi_{\nu]}d{\cal A}$ where $n$ is another
independent normal to the horizon satisfying $n.\xi=1$ on $\cal
H$. Let $\Sigma$ be a spacelike hypersurface extending from $\cal H$
out to spatial infinity. Then, with these conventions, if $X^{\mu\nu}$
is any antisymmetric tensor, Stoke's formula takes the form
\be
2\int_\Sigma dS_\mu\nabla_\nu X^{\mu\nu} = \oint_\infty dS_{\mu\nu}
X^{\mu\nu} - \oint_{\cal H} dS_{\mu\nu} X^{\mu\nu}.
\ee
Applying this to the expressions for $J$ and $J_H$ and making use of
the identity $\nabla_\mu\nabla_\nu
m_\rho=R_{\rho\nu\mu\sigma}m^\sigma$ satisfied by any Killing vector
gives
\be
J - J_H = {1\over 8\pi} \int_\Sigma dS_\mu {R^\mu}_\nu m^\nu.
\label{JJH}
\ee
Thus for vacuum solutions $J_H=J$, the total angular momentum of the
black hole. In general they differ and the difference $J-J_H$ may be
interpreted as the angular momentum of the matter fields present.
Using these formulae it is straightforward to derive Smarr-type
formulae for the mass:
\be
M = {\kappa{\cal A}\over 4\pi} + 2\Omega_H J - {1\over 4\pi}
\int_\Sigma dS_\mu {R^\mu}_\nu \xi^\nu
\label{Smarr2}
\ee
or alternatively:
\be
M = {\kappa{\cal A}\over 4\pi} + 2\Omega_H J_H - {1\over 4\pi}
\int_\Sigma dS_\mu {R^\mu}_\nu k^\nu
\ee
In Einstein-Maxwell theory the integral over $\Sigma$ in
(\ref{Smarr2}) is easily evaluated giving $\Phi_HQ$~~(+$\Psi_HP$ if
the black hole has magnetic charge as well). However, in non-linear
electrodynamics there is no easy way to evaluate the integral and a
simple check for known solutions, such as the static Born-Infeld black
holes discussed in section~6, shows that the simple Smarr formula from
Einstein-Maxwell theory no longer holds.

Now consider a small variation of this solution to another stationary
axi-symmetric solution. Denote the variation of the metric $\delta
g_{\mu\nu}$ by $h_{\mu\nu}$ where the indices on $h$ may be freely
raised and lowered using the metric $g_{\mu\nu}$. Then the variation
of the inverse metric is given by $\delta g^{\mu\nu}=-h^{\mu\nu}$. Now
there is a certain gauge freedom when comparing two different
solutions which may be used to ensure that the stationary and
rotational Killing vectors do not change under the variation:
\be
\delta k^\mu = 0 \quad\hbox{and}\quad \delta m^\mu =0.
\label{deltakm}
\ee
The gauge freedom may also be used to ensure that the horizon $\cal H$
is in the same position in both solutions which is consistent with
(\ref{deltakm}) since the horizon is a fixed point set of $k$ and
$m$. Then the variation of $\xi^\mu$ is given by
\be
\delta\xi^\mu = \delta\Omega_H m^\mu.
\label{deltaxi}
\ee
Since the horizon is in the same position in both solutions the
vectors normal to it in each solution must be parallel. So
$\delta\xi_\mu=f\xi_\mu$ and $\delta n^\mu=gn^\mu$ on $\cal H$ for
some functions $f$ and $g$. Preservation of the normalization
condition $n.\xi=1$ implies that $f+g=0$. This then gives the useful
relation
\be
n^\mu \delta\xi_\nu + \delta n^\mu \xi_\nu = 0 \quad\hbox{on }{\cal H}.
\label{Useful1}
\ee
Another useful relation comes from the fact that the variation
preserves time-translational and rotational invariance so that
\be
{\cal L}_\xi \delta\xi_\mu = \xi^\nu\nabla_\nu \delta\xi_\mu +
\delta\xi_\nu\nabla_\mu\xi^\nu = 0.
\label{Useful2}
\ee
The next step is to calculate the variation of the surface gravity
$\kappa$. The most convenient definition of $\kappa$ to use for this
purpose is
\be
\kappa = -\half n^\mu\nabla_\mu \left( \xi^\nu\xi_\nu \right)
\quad\hbox{on }{\cal H}.
\ee
This then gives
\be
\delta\kappa = -\half \left( n^\mu\xi^\nu + n^\nu\xi^\mu \right)
\nabla_\mu\delta\xi_\nu - \delta\Omega_H n_\mu\xi_\nu \nabla^\mu m^\nu
\quad\hbox{on }{\cal H}
\ee
where use has been made of (\ref{deltaxi}), (\ref{Useful1}),
(\ref{Useful2}) and the fact that the time-translational and
rotational group actions commute $[k,m]=0$. Now using the fact that
$\delta\xi_\mu$ is parallel to $\xi_\mu$, the first term on the right
hand side may be simplified to give
$-\half\nabla^\mu\delta\xi_\mu$. But
$\delta\xi_\mu=h_{\mu\nu}\xi^\nu+\delta\Omega_H m_\mu$ and so the
variation of the surface gravity simplifies giving
\be
\delta\kappa = -\half\xi^\mu\nabla^\nu h_{\mu\nu} - \delta\Omega_H
n_\mu\xi_\nu \nabla^\mu m^\nu \quad\hbox{on }{\cal H}.
\ee
It is now necessary to integrate this over the horizon $\cal H$ in
order to derive the First Law. With some foresight, the first term on
the right hand side of the above expression can be put into the more
convenient form:
\be
-\half\xi^\mu\nabla^\nu h_{\mu\nu} = -\left( n_\mu\xi_\nu -
  n_\nu\xi_\mu \right) \xi^\mu\nabla^{[\lambda} {h^{\nu]}}_\lambda
\quad\hbox{on }{\cal H}
\ee
where use has been made of the fact that the variation is to another
stationary axi-symmetric solution and hence ${\cal L}_\xi h=0$. Thus
\be
\delta\kappa \, d{\cal A} = -\half dS_{\mu\nu} \left(
  2\xi^\mu\nabla^{[\lambda} {h^{\nu]}}_\lambda +
  \delta\Omega_H\nabla^\mu m^\nu \right) \quad\hbox{on }{\cal H}
\ee
and integrating over $\cal H$ gives
\be
{\cal A}\delta\kappa = -8\pi\delta\Omega_HJ_H - \oint_{\cal H}
dS_{\mu\nu} \xi^\mu\nabla^{[\lambda} {h^{\nu]}}_\lambda
\label{30}
\ee
since $\kappa$ is constant over $\cal H$ by the Zeroth Law. Using
Stoke's formula this last integral can be converted to an integral at
spatial infinity and an integral over $\Sigma$. The integral over
spatial infinity may be evaluated using the asymptotic form of the
metric:
\be
\oint_\infty dS_{\mu\nu} \xi^\mu\nabla^{[\lambda} {h^{\nu]}}_\lambda =
4\pi\delta M.
\ee
So equation (\ref{30}) becomes
\be
4\pi\delta M + {\cal A}\delta\kappa + 8\pi\delta\Omega_HJ_H =
\int_\Sigma dS_\mu \xi^\mu\nabla_\nu\nabla^{[\lambda}
{h^{\nu]}}_\lambda
\label{32}
\ee
where again use has been made of the fact that ${\cal L}_\xi
h_{\mu\nu}=0$ in rearranging the final integral. Now the integrand in
this expression is related to the variation of the Ricci scalar $R$
under metric perturbations:
\be
\delta R = 2\nabla_\nu\nabla^{[\lambda} {h^{\nu]}}_\lambda -
h^{\mu\nu}R_{\mu\nu}.
\ee
So using this and also making use of (\ref{JJH}) and (\ref{deltaxi})
to eliminate $J_H$ in favour of $J$ in (\ref{32}) gives
\be
4\pi\delta M + {\cal A}\delta\kappa + 8\pi\delta\Omega_HJ =
\int_\Sigma dS_\mu \left\{ \half\xi^\mu \left( \delta R +
h^{\rho\sigma}R_{\rho\sigma} \right) + {R^\mu}_\nu\delta\xi^\nu
\right\}.
\label{34}
\ee
Using the Einstein equations and the fact that $\delta dS_\mu=\half
h\,dS_\mu$ where $h=h_{\mu\nu}g^{\mu\nu}$ equation (\ref{34}) may be
rearranged to give
\be
4\pi\delta M + {\cal A}\delta\kappa + 8\pi\delta\Omega_HJ =
\int_\Sigma dS_\mu \left( \xi^\mu T_{\rho\sigma}h^{\rho\sigma} +
  2{T^\mu}_\nu\delta\xi^\nu \right) + {\half}\delta \int_\Sigma
dS_\mu \xi^\mu R.
\label{36}
\ee
Finally, combining this with the variation of the Smarr formula
(\ref{Smarr2}) and using the Einstein equations again gives the First
Law of Black Hole Mechanics for stationary axi-symmetric black holes
in General Relativity coupled to arbitrary matter fields with
energy-momentum tensor $T_{\mu\nu}$:
$$
\delta M = {\kappa\over 8\pi}\delta{\cal A} + \Omega_H\delta J +
{1\over 8\pi}\int_\Sigma dS_\mu \xi^\mu T_{\rho\sigma}h^{\rho\sigma} -
{1\over 4\pi}\delta\int_\Sigma dS_\mu {T^\mu}_\nu\xi^\nu
$$
\be
+
{1\over 4\pi}\delta\Omega_H\int_\Sigma dS_\mu {T^\mu}_\nu m^\nu.
\label{39}
\ee
In the case where the matter fields may be described by a Lagrangian
$L_F$, the energy-momentum tensor $T_{\mu\nu}$ is related to the
variation of the Lagrangian with respect to the metric:
\be
\delta_g \left( \sqrt{g}L_F \right) = 2\sqrt{g}T_{\mu\nu}h^{\mu\nu}.
\ee
Here $\delta_g$ denotes variation with respect to $g_{\mu\nu}$ and
$\delta_A$ will denote variation with respect to the matter
fields and so $\delta=\delta_g+\delta_A$. Using these relations the
First Law may be rewritten in the form
$$
\delta M = {\kappa\over 8\pi}\delta{\cal A} + \Omega_H\delta J -
{1\over 4\pi}\delta\int_\Sigma dS_\mu \left( {T^\mu}_\nu\xi^\nu -
  \fraction{1}{4}L_F\xi^\mu \right) - {1\over
  16\pi}\delta_A\int_\Sigma dS_\mu \xi^\mu L_F
$$
\be
+
{1\over 4\pi}\delta\Omega_H\int_\Sigma dS_\mu \left(
  {T^\mu}_\nu m^\nu - \fraction{1}{4}L_F m^\mu \right).
\label{FLLag}
\ee
Note that by using equation (\ref{JJH}), this statement of the First
Law may be written in terms of $\delta J_H$ rather than $\delta J$. In
this case $\xi$ is replaced by $k$ in the first two integrals of
(\ref{FLLag}) and the final integral is absent. This then agrees with
equation (90) of \cite{HeuStr}. However, in going from equation (90)
to (92) in \cite{HeuStr}, the variation of $\Omega_H$ was omitted.
Thus the final term in (\ref{FLLag}) corrects equation (92) of
\cite{HeuStr}, although their derivation up to that point remains
correct. For the purposes of this paper, the form of the First Law
(\ref{FLLag}), in terms of $\xi$ and $\delta J$, is more convenient
because the electric and magnetic potentials $\Phi$ and $\Psi$ must be
defined in terms of $\xi$ (not $k$) so that they are constant on the
horizon.

The last term in equation (\ref{FLLag}) has an interesting
interpretation. It may be regarded as the contribution to the mass
variation due to the angular momentum of the matter fields outside the
event horizon. For example, if the surface $\Sigma$ is chosen so that
it is invariant under the rotational symmetry, i.e.\ $dS_\mu m^\mu=0$,
then this last term evaluates to $(J-J_H)\delta\Omega_H$. Note that in
all the calculations above, no specific assumptions have been made
about the surface $\Sigma$ other than the fact that it has a boundary
on $\cal H$ and extends out to spatial infinity. Since, in general,
the integrands in the above formula are not total derivatives, the
integrals will depend on the choice of $\Sigma$.

In the next section the case of non-linear electrodynamics will be
discussed. In this case the integrand of the final integral in
(\ref{FLLag}) is a total derivative and so the integral is independent
of the particular choice of $\Sigma$. So $\Sigma$ may indeed be chosen
to be invariant under the rotational symmetry, to simplify the
evaluation of the integral.  Thus, in this case, the final integral
gives the angular momentum of the electromagnetic field, $J_F=J-J_H$
and moreover this result is independent of the choice of $\Sigma$.


\sect{Application to Non-Linear Electrodynamics}

Using the special form of the energy-momentum tensor for non-linear
electrodynamics (\ref{EnMom}) the First Law may be written as
\be
\delta M = {\kappa\over 8\pi}\delta{\cal A} + \Omega_H\delta J -
\delta I_1 - \delta_A I_2 + \delta\Omega_H I_3
\label{FL123}
\ee
where
\ba
 & & I_1 = {1\over 4\pi} \int_\Sigma dS_\mu
 G^{\mu\lambda}F_{\nu\lambda}\xi^\nu, \nonumber \\
 & & \nonumber \\
 & & I_2 = {1\over 16\pi} \int_\Sigma dS_\mu \xi^\mu L_F, \\
 & & \nonumber \\
 & & I_3 = {1\over 4\pi} \int_\Sigma dS_\mu
 G^{\mu\lambda}F_{\nu\lambda} m^\nu. \nonumber
\ea
Using (\ref{EH}) and (\ref{EPhi}) together with the electromagnetic
equations of motion (\ref{EOM}) enables the integrand in $I_1$ to be
written as a total derivative:
\be
I_1 = -{1\over 4\pi} \int_\Sigma dS_\mu\nabla_\nu \left( \Phi
  G^{\mu\nu} \right).
\ee
Using Stoke's formula and the asymptotic condition $\Phi\rightarrow 0$
as $r\rightarrow\infty$, $I_1$ may be expressed as an integral over
the horizon. Now the electric charge of the black hole is given by
\be
Q = -{1\over 8\pi} \oint dS_{\mu\nu} G^{\mu\nu}
\ee
where the integral may be taken over any closed 2-surface enclosing
the charge and is independent of the particular surface chosen by
virtue of the equations of motion (\ref{EOM}). Therefore, using the
fact that $\Phi$ is constant on the horizon, one finds that
\be
I_1 = -\Phi_HQ.
\ee
Evaluation of the second integral $I_2$ is not possible, in general,
since the precise form of $L_F$ is not known. However, in the
calculation of the First Law it is not necessary to evaluate $I_2$,
only its variation with respect to $A_\mu$. This is done with the help
of the definition of $G^{\mu\nu}$ in terms of the variation of $L_F$
with respect to $F_{\mu\nu}$ which gives
\be
\delta_A L_F = {\partial L_F\over\partial F_{\mu\nu}} \delta_A
F_{\mu\nu} = -2 G^{\mu\nu}\delta_AF_{\mu\nu}
\ee
and so
\be
\delta_A I_2 = -{1\over 8\pi} \int_\Sigma dS_\mu \xi^\mu
G^{\rho\sigma}\delta_AF_{\rho\sigma}.
\label{74}
\ee
In order to express the integrand in (\ref{74}) in a more convenient
form, consider the vector
\be
Y^\mu = \epsilon^{\mu\nu\rho\sigma} \left( \delta_A E_\nu \star
G_{\rho\sigma} - H_\nu \delta_A F_{\rho\sigma} \right).
\ee
By using the definitions of $E_\mu$ and $H_\mu$ (\ref{EH}) and
expanding out products of $\epsilon$ tensors, $Y^\mu$ becomes
\be
Y^\mu = \xi^\mu G^{\rho\sigma}\delta_AF_{\rho\sigma}
\ee
which is precisely the integrand in (\ref{74}). Alternatively, using
$\star\star=-1$, $Y^\mu$ may be written as
\be
Y^\mu = -2 \left( \delta_A E_\nu G^{\mu\nu} + H_\nu \delta_A \star
F^{\mu\nu} \right).
\ee
Now from equations (\ref{EH}) and (\ref{deltaxi}) the variation of
$E_\nu$ is
\be
\delta_A E_\nu = \left( \delta-\delta_g \right) E_\nu = \delta E_\nu -
\delta\Omega_H F_{\nu\lambda}m^\lambda
\ee
and so
\ba
Y^\mu &=& -2 \left( \delta E_\nu G^{\mu\nu} + H_\nu \delta_A\star
F^{\mu\nu} \right) + 2\delta\Omega_H G^{\mu\nu}F_{\nu\lambda}m^\lambda
\nonumber \\
 &=& -2\nabla_\nu \left( \delta\Phi G^{\mu\nu} + \Psi\delta_A\star
   F^{\mu\nu} \right) + 2\delta\Omega_H
 G^{\mu\nu}F_{\nu\lambda}m^\lambda
\ea
using (\ref{EPhi}), (\ref{HPsi}), the Bianchi identities
(\ref{Bianchi}) and the equations of motion (\ref{EOM}) to obtain the
last line. Thus using Stoke's formula and the asymptotic conditions
$\Phi,\Psi\rightarrow 0$ as $r\rightarrow\infty$ the variation of
$I_2$ may be written as an integral over the horizon plus an integral
over $\Sigma$ which can be seen to be proportional to the third
integral $I_3$:
\be
\delta_A I_2 = -{1\over 8\pi}\oint_{\cal H} dS_{\mu\nu} \left(
  \delta\Phi G^{\mu\nu} + \Psi\delta_A\star F^{\mu\nu} \right) +
\delta\Omega_H I_3.
\ee
Now defining the magnetic charge $P$ as
\be
P = {1\over 8\pi} \oint dS_{\mu\nu} \star F^{\mu\nu}
\ee
and using the fact that $\Phi$ and $\Psi$ are constant on the horizon,
the variation of $I_2$ may be evaluated giving
\be
\delta_A I_2 = Q\delta\Phi_H - \Psi_H\delta P + \delta\Omega_H I_3.
\ee
Substituting this into equation (\ref{FL123}) shows that there is now
no need to evaluate the final integral $I_3$ since it cancels in
(\ref{FL123}) and the First Law of Black Hole Mechanics for stationary
axi-symmetric black holes in non-linear electrodynamics has been
established:
\be
\delta M = {\kappa\over 8\pi}\delta{\cal A} + \Omega_H\delta J +
\Phi_H\delta Q + \Psi_H\delta P.
\label{NLEFL}
\ee
This agrees with the known formula for Kerr-Newman black holes in
Einstein-Maxwell theory, which is of course a special case of the
above calculation.

It is interesting to note that the final term $\delta\Omega_HI_3$ in
(\ref{FL123}), which would have given a contribution
$(J-J_H)\delta\Omega_H$ due to the angular momentum of the
electromagnetic field, cancelled with part of the previous term
$\delta_AI_2$. This is a result of the fact that {\em co-rotating}
electric and magnetic fields were used in the calculation (in order
that their potentials be constant on the horizon). Thus the rotational
effects of the electromagnetic field are automatically taken into
account in the last two terms of (\ref{NLEFL}).


\sect{Born-Infeld Theory}

The one non-linear theory of electrodynamics which keeps making
appearances again and again in many different contexts within modern
theoretical physics is Born-Infeld theory. Amongst its many special
properties is an exact $SO(2)$ electric-magnetic duality
invariance. The Lagrangian density describing Born-Infeld theory (in
arbitrary spacetime dimensions) is
\be
{\goth L}_F = \sqrt{g}L_F = {4\over b^2} \left\{ \sqrt{g} -
  \sqrt{|\det(g_{\mu\nu}+bF_{\mu\nu})|} \right\}
\label{LBI1}
\ee
where $b$ is a fundamental parameter of the theory, with dimensions of
mass. In open superstring theory, for example, loop calculations lead
to this Lagrangian with $b=2\pi\alpha'$. In four spacetime dimensions
the determinant in (\ref{LBI1}) may be expanded out to give
\be
L_F = {4\over b^2} \left\{ 1 - \sqrt{1+\half
    b^2F^2-\fraction{1}{16}b^4(F\star F)^2} \right\}
\label{LBI2}
\ee
which coincides with the usual Maxwell Lagrangian in the weak field
limit.

The tensor $G^{\mu\nu}$ defined in section~2 is given by
\be
G^{\mu\nu} = -{1\over 2} {\partial L_F\over\partial F_{\mu\nu}} =
{F^{\mu\nu} - \fraction{1}{4}b^2(F\star F)\star F^{\mu\nu} \over
  \sqrt{1+\half b^2F^2-\fraction{1}{16}b^4(F\star F)^2}}
\label{GBI}
\ee
(so that $G^{\mu\nu}\approx F^{\mu\nu}$ for weak fields) and satisfies
the electromagnetic equations of motion
\be
\nabla_\mu G^{\mu\nu} = 0
\label{EOMBI}
\ee
which are highly non-linear in $F_{\mu\nu}$. The energy-momentum
tensor may be written as
\be
T_{\mu\nu} = {{F_\mu}^\lambda F_{\nu\lambda} +
  \fraction{1}{b^2}\left[\sqrt{1+\half
      b^2F^2-\fraction{1}{16}b^4(F\star F)^2} -1 -\half
    b^2F^2\right]g_{\mu\nu} \over \sqrt{1+\half
    b^2F^2-\fraction{1}{16}b^4(F\star F)^2}}.
\label{EnMomBI}
\ee
Although it is by no means obvious, it may be verified that equations
(\ref{GBI})--(\ref{EnMomBI}) are invariant under electric-magnetic
duality: $F\leftrightarrow\star G$.

In flat space, and for purely electric configurations, the Lagrangian
(\ref{LBI2}) reduces to
\be
L_F = {4\over b^2} \left\{ 1 - \sqrt{1-b^2\mbox{\boldmath $E$}^2}
\right\}
\ee
so there is an upper bound on the electric field strength
$\mbox{\boldmath $E$}$:
\be
\left|\mbox{\boldmath $E$}\right| \le {1\over b}.
\label{Ebound}
\ee
The field due to a point charge is
\be
E_r = {Q\over\sqrt{r^4+b^2Q^2}}
\ee
and so achieves the bound (\ref{Ebound}) at $r=0$. The total
self-energy of the point charge is thus
\be
{\cal E} = {1\over 4\pi} \int d^3\mbox{\boldmath $x$} T_{00} = {1\over
  4\pi} \int d^3\mbox{\boldmath $x$} {1\over
  b^2r^2}\left(\sqrt{r^4+b^2Q^2}-r^2\right).
\ee
Integrating by parts gives a standard elliptic integral:
\be
{\cal E} = {2Q^2\over 3} \int_0^\infty {dr\over\sqrt{r^4+b^2Q^2}} =
{(\pi Q)^{3\over 2}\over 3\sqrt{b}\,\Gamma\!\left({3\over 4}\right)^2}
\ee
which is finite (for simplicity, $Q$ and $b$ are taken to be positive
here). Thus Born-Infeld theory succeeded in its original goal of
providing a model for point charges with finite self-energy. Note that
in the limit $b\rightarrow 0$, Maxwell theory is reproduced and the
self-energy diverges.

Now consider static spherically symmetric black holes in this
theory. Using electric-magnetic duality, there is no loss of
generality in considering only electrically charged black holes. The
solution is
$$
ds^2 = -\left(1-{2m(r)\over r}\right)dt^2 + \left(1-{2m(r)\over
    r}\right)^{-1}dr^2 + r^2\left(d\theta^2+\sin^2\theta
  d\phi^2\right),
$$
\be
G_{tr} = {Q\over r^2} \quad\hbox{or}\quad F_{tr} =
 {Q\over\sqrt{r^4+b^2Q^2}}, \nonumber
\ee
where the function $m(r)$ satisfies
\be
m'(r) = {1\over b^2} \left(\sqrt{r^4+b^2Q^2}-r^2\right)
\label{mprimed}
\ee
and $'$ denotes differentiation with respect to $r$. The mass $M$ is
given by
\be
M = \lim_{r\rightarrow\infty} m(r)
\ee
and hence
\be
m(r) = M - {1\over b^2}\int_r^\infty dx
\left(\sqrt{x^4+b^2Q^2}-x^2\right),
\ee
which is a monotonically increasing function of $r$. The horizons are
given by the roots of the equation $r=2m(r)$ and so the number of
horizons will be determined by $m(0)$ and $m'(0)$. $m(0)$ depends on
the self-energy of the electromagnetic field, the integral being the
same as for the point charge in flat space:
\be
m(0) = M - {(\pi Q)^{3\over 2}\over 3\sqrt{b}\,\Gamma\!\left({3\over
      4}\right)^2} = M - {\cal E}
\ee
and so $m(0)$ may be interpreted as the binding energy. From
(\ref{mprimed}) one has
\be
m'(0) = {Q\over b}.
\ee
For $m(0)>0$ there is precisely one non-degenerate horizon. If
$m(0)=0$ then there is one non-degenerate horizon for $Q>\half b$ and
none otherwise. The case $m(0)<0$ is similar to Reissner-Nordstr\"om,
with either no horizons, one degenerate horizon or two non-degenerate
horizons, depending on the relative magnitudes of $M$, $Q$ and $b$.
Note that the Reissner-Nordstr\"om solution is recovered in the limit
$b\rightarrow 0$ in which case $m(0)\rightarrow-\infty$.

Assuming that there is at least one horizon, let $r_+$ denote the
outer event horizon. It has surface gravity $\kappa$ given by
\be
\kappa = -{1\over 2} \left. {dg_{tt}\over dr} \right|_{r=r_+} =
{1\over 2r_+} - {1\over b^2r_+} \left(\sqrt{r_+^4+b^2Q^2}-r_+^2\right)
\ee
and a surface area $\cal A$ of
\be
{\cal A} = 4\pi r_+^2.
\ee
The electric potential on the horizon (in a gauge in which it vanishes
at infinity) is
\be
\Phi_H = \int_{r_+}^\infty dr {Q\over\sqrt{r^4+b^2Q^2}}
\ee
and it is simple to verify that the First Law
\be
dM = {\kappa\over 8\pi}d{\cal A} + \Phi_HdQ
\ee
is indeed satisfied. On the other hand, the usual statement of Smarr's
formula (\ref{Smarr}) does not hold.


\sect{Inclusion of Scalar Fields}

In all the cases in which a non-linear theory of electrodynamics
arises from string theory or D-brane theories, there are always
additional scalar fields present. In general, these fields are coupled
to the electromagnetic field in some non-trivial way. It is thus
interesting to ask in what ways the presence of scalars affects the
results obtained above.  The simplest case would be to consider a
single scalar field. For example, Einstein-Maxwell-dilaton theory
which may be described by the Lagrangian
\be
L_F = -2\left( \nabla\phi \right)^2 - e^{-2\alpha\phi}F^2.
\label{EMDLag}
\ee
In the case $\alpha=1$ this coincides with a certain limit of string
theory and $\alpha=-\sqrt{3}$ gives Kaluza-Klein theory. One may also
include an axion. For example, the Lagrangian
\be
L_F = -2\left( \nabla\phi \right)^2 -2e^{4\phi}\left( \nabla a
\right)^2 -e^{-2\phi}F^2 + 2aF\star F
\label{EMDALag}
\ee
describes the bosonic sector of $N=4$ supergravity in 4 dimensions and
also string theory compactified on a torus. The equations of motion
derived from the above Lagrangian have an $SL(2,{\Bbb R})$ invariance
which includes simple electric-magnetic duality as a subgroup. Of
course, there may also be higher order corrections giving non-linear
electrodynamics coupled to an axion and a dilaton. In this case, under
certain circumstances, the theory may retain its $SL(2,{\Bbb R})$
duality \cite{GibRas2}. This is thus of interest in string theory
where the duality is S-duality, relating the strong and weak coupling
limits of the theory.

To try to extend the results of this paper to include theories of the
above type, consider the Lagrangian
\be
L_F = -2 {\cal G}_{ab}(\phi^c) (\nabla_\mu\phi^a)(\nabla^\mu\phi^b) +
\widetilde{L}_F(F_{\mu\nu},\phi^a).
\label{ScalarLag}
\ee
Here ${\cal G}_{ab}$ is the (symmetric) metric on the target space of
the scalar fields and $\widetilde{L}_F$ is a function of the scalars
$\phi^a$ as well as $F_{\mu\nu}$ but contains no derivative terms.
Clearly then, the two Lagrangians (\ref{EMDLag}) and (\ref{EMDALag})
above are special cases of this one. The tensor $G^{\mu\nu}$ is
defined in the same way as in section~2:
\be
G^{\mu\nu} = -{1\over 2} {\partial L_F\over\partial F_{\mu\nu}} =
-{1\over 2} {\partial\widetilde{L}_F\over\partial F_{\mu\nu}}
\ee
and so now it will, in general, depend on the scalar fields as
well as $F_{\mu\nu}$. The electromagnetic equations of motion,
obtained by varying with respect to $A_\mu$, are again
\be
\nabla_\mu G^{\mu\nu} = 0.
\ee
So the conserved electric and magnetic charges are defined as before.
Varying the action with respect to $\phi^a$ gives the scalar equations
of motion:
\be
-4\nabla^\mu\left( {\cal G}_{ab}\nabla_\mu\phi^b \right) =
{\partial\widetilde{L}_F\over\partial\phi^a} - 2{\partial{\cal
    G}_{bc}\over\partial\phi^a} (\nabla_\mu\phi^b)(\nabla^\mu\phi^c).
\ee
Varying with respect to the metric gives the Einstein field equations
and the energy-momentum tensor now contains kinetic terms coming from
the scalar fields:
\be
T_{\mu\nu} = {G_\mu}^\lambda F_{\nu\lambda} + {\cal
  G}_{ab}(\nabla_\mu\phi^a)(\nabla_\nu\phi^b) +
\fraction{1}{4}g_{\mu\nu}L_F.
\ee

Considering stationary axi-symmetric solutions, the scalar fields must
be time-translationally and rotationally invariant,
${\cal L}_k\phi^a=0$ and ${\cal L}_m\phi^a=0$. Hence
\be
{\cal L}_\xi\phi^a = \xi^\mu\partial_\mu\phi^a = 0.
\label{Liephi}
\ee
Therefore, contractions of $T_{\mu\nu}$ with $\xi^\mu$ are unaltered
in form by the presence of the scalar fields and so the proofs that
$\Phi_H$, $\Psi_H$ and $\kappa$ are constant on the horizon $\cal H$
follow as before. Also, for the same reason, the evaluation of the
first and third integrals in (\ref{FLLag}) is unchanged. Thus the only
change to the First Law comes from the second integral. This will
contribute an additional term
\be
-{1\over 16\pi}\delta_\phi\int_\Sigma dS_\mu \xi^\mu L_F
\label{Additional}
\ee
where $\delta_\phi$ denotes the variation with respect to the scalar
fields.

Using the scalar equations of motion, the variation of the Lagrangian
$L_F$ with respect to $\phi^a$ may be written as a total derivative:
\ba
\delta_\phi L_F &=& -4{\cal
  G}_{ab}(\nabla_\mu\phi^a)(\nabla^\mu\delta\phi^b) + \left[
  {\partial\widetilde{L}_F\over\partial\phi^a} - 2{\partial{\cal
      G}_{bc}\over\partial\phi^a}(\nabla_\mu\phi^b)(\nabla^\mu\phi^c)
\right] \delta\phi^a \nonumber \\
 & & \nonumber \\
 &=& -4{\cal G}_{ab}(\nabla_\mu\phi^a)(\nabla^\mu\delta\phi^b)
 -4\nabla^\mu\left({\cal G}_{ab}\nabla_\mu\phi^b\right)\delta\phi^a \\
 & & \nonumber \\
 &=& -4\nabla^\mu\left({\cal
     G}_{ab}(\nabla_\mu\phi^a)\delta\phi^b\right). \nonumber
\ea
Thus the variation of the integrand in (\ref{Additional}) with respect
to $\phi$ is also a total derivative:
\be
\xi^\mu\delta_\phi L_F = -8\nabla_\nu\left( \xi^{[\mu}{\cal
    G}_{ab}(\nabla^{\nu]}\phi^a)\delta\phi^b \right)
\ee
where use has been made of the fact that ${\cal L}_\xi({\cal
  G}_{ab}\nabla^\mu\phi^a\delta\phi^b)=0$. So, using Stoke's formula,
the additional contribution (\ref{Additional}) to the First Law may be
written as surface integrals over the horizon $\cal H$ and at spatial
infinity. The contribution from the horizon vanishes identically using
(\ref{Liephi}). Therefore the only additional contribution is an
integral over spatial infinity:
\be
{1\over 4\pi} \oint_\infty dS_{\mu\nu} \xi^\mu {\cal G}_{ab}
(\nabla^\nu\phi^a) \delta\phi^b.
\ee
This integral is easily evaluated using the asymptotic form of the
scalar fields at infinity
\be
\phi^a \sim \phi^a_\infty + {\Sigma^a\over r} + {\cal O}(r^{-2})
\ee
where $\Sigma^a$ are the scalar charges. The integral then gives
${\cal G}_{ab}\Sigma^a\delta\phi^b_\infty$ and so the modified First
Law in the presence of scalar fields is
\be
\delta M = {\kappa\over 8\pi}\delta{\cal A} + \Omega_H\delta J +
\Phi_H\delta Q + \Psi_H\delta P + {\cal
  G}_{ab}\Sigma^a\delta\phi^b_\infty.
\label{FLScalar}
\ee
This agrees with the result obtained in \cite{GibKalKol}. In that case
allowing the asymptotic values of the scalar fields to vary was of
relevance to string theory where different asymptotic values label
different vacua of the theory.

Of course one may also consider more than one $U(1)$ gauge field and
the above result generalizes in the obvious way. This is relevant to
the bosonic sector of supergravity theories in 4 dimensions for
example and also string theories compactified down to 4 dimensions. In
fact the Lagrangian (\ref{ScalarLag}) is general enough to cover a
very wide range of theories. The statement of the First Law
(\ref{FLScalar}) is consistent with the ``no hair'' theorems,
indicating that the black holes of such theories are completely
specified by their mass, angular momentum, electric and magnetic
charges and the asymptotic values of the scalars. The scalar charges
are then determined uniquely.

It was remarked above that, because the First Law does hold for
non-linear electrodynamics, the Smarr formula does not. To make this
more explicit, consider first the case of {\em linear}
electrodynamics. In this case the Lagrangian must be quadratic in
$F_{\mu\nu}$. So, considering $N$ $U(1)$ gauge fields labelled by
$\Lambda,\Sigma=1,2,\dots N$, the Lagrangian must take the form
\be
L_F = -2 {\cal G}_{ab} (\nabla_\mu\phi^a)(\nabla^\mu\phi^b) -
\mu_{\Lambda\Sigma}F^\Lambda_{\mu\nu}F^{\Sigma\mu\nu} +
\nu_{\Lambda\Sigma}F^\Lambda_{\mu\nu}\star F^{\Sigma\mu\nu}
\ee
where $\mu_{\Lambda\Sigma}$ and $\nu_{\Lambda\Sigma}$ are arbitrary
functions of the scalar fields $\phi^a$. This is precisely the class
of theories discussed in \cite{GibKalKol}. Now, using the no hair
theorem, the mass may be regarded as a function of the horizon area
$\cal A$, the angular momentum $J$, the electric and magnetic charges
$Q^\Lambda$, $P^\Lambda$ and the asymptotic values of the scalar
fields $\phi^a_\infty$. Furthermore, on dimensional grounds, it is
clear that any other parameters of the theory must be
dimensionless\footnote{recall $G=c=4\pi\epsilon_0=1$ in this paper, so
  the dimensions of any quantity may be expressed in terms of mass
  dimensions} and so $M$ is a homogeneous function. Using the fact
that $\cal A$ and $J$ have units of $(\hbox{mass})^2$, $Q^\Lambda$ and
$P^\Lambda$ have units of mass and $\phi^a_\infty$ is dimensionless,
Euler's theorem for homogeneous functions implies that
\be
M = 2{\cal A}{\partial M\over\partial\cal A} + 2J{\partial
  M\over\partial J} + Q^\Lambda{\partial M\over\partial Q^\Lambda} +
P^\Lambda{\partial M\over\partial P^\Lambda}.
\ee
The partial derivatives are given by the First Law (\ref{FLScalar})
and thus one obtains Smarr's formula
\be
M = {\kappa{\cal A}\over 4\pi} + 2\Omega_HJ + \Phi_\Lambda Q^\Lambda +
\Psi_\Lambda P^\Lambda.
\ee
Note that its form is unaltered by the presence of the scalar fields
as was pointed out in \cite{GibKalKol}.

Now consider non-linear theories of electrodynamics. For example,
consider the higher order corrections which result from loop
calculations in string theory, leading to Born-Infeld-type
Lagrangians. Since the Lagrangian is no longer quadratic in
$F_{\mu\nu}$, one may argue on dimensional grounds that there will
always be a coupling constant with non-trivial mass dimensions (for
example, the constant $b$ in Born-Infeld theory discussed in the last
section). In general this coupling constant will enter into the
formula for the mass and so $M$ will no longer be a homogeneous
function of $\cal A$, $J$, $Q^\Lambda$, $P^\Lambda$ and
$\phi^a_\infty$. Thus Smarr's formula will no longer hold but, as one
might have expected, the First Law remains true, i.e.\ the black
hole's thermodynamics are preserved by higher order quantum
corrections.


\bigskip
\noindent
\hfil{\large\bf Acknowledgements}\hfil

\medskip

I would like to thank Gary Gibbons for many stimulating discussions
and for helpful comments and corrections during the preparation of
this paper. This work was supported by EPSRC grant no.\ 9400616X.

\end{document}